\documentclass[conference]{IEEEtran}
\IEEEoverridecommandlockouts
  \usepackage{cite}
  \usepackage{amsmath,amssymb,amsfonts}
  \usepackage{algorithmic}
  \usepackage{graphicx}
  \usepackage{textcomp}
  \usepackage{xcolor}
  
  \usepackage{tikz}
  \usetikzlibrary{shapes,arrows,quotes}
  \usepackage{pgfplots}
  \pgfplotsset{compat=1.18, width=7.5cm}
   
  \def\BibTeX{{\rm B\kern-.05em{\sc i\kern-.025em b}\kern-.08em
      T\kern-.1667em\lower.7ex\hbox{E}\kern-.125emX}}
  \usepackage{tikz}
  \usetikzlibrary{arrows,fit,positioning,shapes,patterns}
  \usepackage{subcaption}
  \usepackage{listings}
  \usepackage{soul}
  \newcommand{\egg}{\texttt{egg}}

\begin{document}

  \tikzstyle{block} = [draw, rectangle, minimum height=3em, minimum width=6em]
  
  \title{Datapath Verification via Word-Level E-Graph Rewriting}
  
  \author{\IEEEauthorblockN{Samuel Coward$^{1,2}$, Emiliano Morini$^1$, Bryan Tan$^1$, Theo Drane$^1$ and George A.~Constantinides$^2$}
  \IEEEauthorblockA{$^1$ Intel Corporation, $^2$ Imperial College London,\\
  Email: \{samuel.coward, emiliano.morini, bryan.tan, theo.drane\}@intel.com, g.constantinides@imperial.ac.uk}
  }
  
  
  \maketitle
  
  \begin{abstract}
  Formal verification of datapath circuits is challenging as they are subject to intense optimization effort in the design phase. Industrial vendors and design companies deploy equivalence checking against a golden or existing reference design to satisfy correctness concerns. State-of-the-art datapath equivalence checking tools deploy a suite of techniques, including rewriting. We propose a rewriting framework deploying bitwidth dependent rewrites based on the e-graph data structure, providing a powerful assistant to existing tools. The e-graph can generate a path of rewrites between the reference and implementation designs that can be checked by a trusted industry tool. We will demonstrate how the intermediate proofs generated by the assistant enable convergence in a state of the art tool, without which the industrial tool runs for 24 hours without making progress. The intermediate proofs automatically introduced by the assistant also reduce the total proof runtime by up to 6x.
  \end{abstract}
  
  
  \section{Introduction}
  Arithmetic datapath circuits like adders and multipliers are included in almost every electronic device and designers implement low-level optimizations targeting the best power, performance and area. As a result, the verification of datapath circuits is challenging since the code can be difficult to review, making it hard to identify a sufficient test suite. Typically, exhaustive simulation is infeasible due to the size of the input space. Undetected bugs lead to system failures and reputation damage~\cite{Pratt1995AnatomyBug}. Formal Verification (FV) is the only scalable option to prove the absence of bugs in hardware~\cite{Darbari2019SmartVerification}.
  
  One of the most successful FV approaches to verify datapath circuit designs is based on Equivalence Checking (EC), where the design under test, usually called the \emph{implementation}, is proven to be equivalent to a golden reference design, often called the \emph{specification}. Electronic Design Automation (EDA) vendors have developed commercial tools drastically lowering the entry barrier \cite{Koelbl2009SolverChecking}, allowing semiconductor companies to fully verify many different designs~\cite{Xue2013SimplificationModels,Morini2019FormalDivision}. 
  
  Commercial tools orchestrate a suite of solver technologies~\cite{Koelbl2009SolverChecking}, including SAT, SMT and BDD based solvers. Yet still some simple designs can not be proven equivalent. For example, an industrial state-of-the-art tool is unable to prove the equivalence of the two designs shown in Figure~\ref{fig:motivational_verilog} without requiring manual effort to apply advanced formal techniques. We enhance the capabilities of such tools by deploying word-level rewriting in combination with a data structure, known as an e(quivalence)-graph. E-graphs are found at the heart of modern SMT solvers~\cite{DeMoura2008Z3:Solver}, but by applying them at the abstraction level used by humans in RTL design we can tailor the rewrites to datapath verification. 
  
  
  \begin{figure}
      \centering
      \begin{subfigure}{.45\columnwidth}
      \input{motivational_verilog_orig}
      \caption{Specification design.}
      \label{fig:orig_verilog}
      \end{subfigure}\quad%
  \begin{subfigure}{.45\columnwidth}
    \centering
      \input{motivational_verilog_opt}
  
      \caption{Implementation design.}
      \label{fig:opt_verilog}
  \end{subfigure}
  \caption{A motivational example, where existing EC tools fail to prove the equivalence of these two designs.}
  \label{fig:motivational_verilog}
  \end{figure}
  
  In this work we modify an existing e-graph based RTL optimization tool~\cite{Coward2022AutomaticE-Graphs} to produce a powerful formal verification assistant. The proposed verification assistant is able to both exceed the capabilities of the industrial state of the art, reduce verification runtimes and decrease the complexity of the EC problem. The approach taken here is similar to that of Stepp, Tate and Lerner, who initially developed an e-graph based LLVM optimizer~\cite{Tate2009EqualityOptimization}, and later modified it to perform translation validation~\cite{Stepp2011Equality-basedLLVM}. We differ from this previous work in that we validate numerically intense optimizations at a lower abstraction level often performed by a human rather than a compiler. We also deploy modern e-graph developments allowing us to incorporate value range analysis techniques and can generate a simplified EC problem for FV engineers. The approach presented is sound, as we check each intermediate step using a trusted EC tool.
  The paper contains the following novel contributions:
  \begin{itemize}
      \item a word-level e-graph framework that composes a set of sub-problems from local rewrites to assist FV tools,
      \item a specialized and extensible bitwidth dependent rewrite set for datapath verification,
      \item an e-graph extraction method minimizing the `distance' between two designs,
      \item test cases showing an enhancement in capabilities over industrial tools, reducing the need for manual FV effort. 
  \end{itemize}
  
  First, we provide the necessary background on verification and e-graphs. In Section \ref{sec:proving_equiv} we describe how word-level e-graphs can be applied to produce a verification assistant. In Section \ref{sec:case_study} we describe a case study where we outperform the industrial state of the art. Lastly, in Section \ref{sec:results} we present results demonstrating overall verification runtime improvements. 
  
  \section{Background}\label{sect:background}
  \subsection{Datapath Verification}
  Classic formal property verification methods successfully used to verify state machines and communication protocols are not able to verify datapath dominated circuits. Theorem Proving \cite{Hunt2017IndustrialACL2, Moore1998AProgram, Russinoff1998Processor, Harrison1999AArithmetic,Harrison2010HandbookReasoning} and Symbolic Trajectory Evaluation~\cite{Seger1995FormalTrajectories} are valuable approaches, but their common downsides are a high barrier to entry and maintenance of complex code bases.
  
\begin{figure}
      \centering
      \includegraphics[scale=0.45]{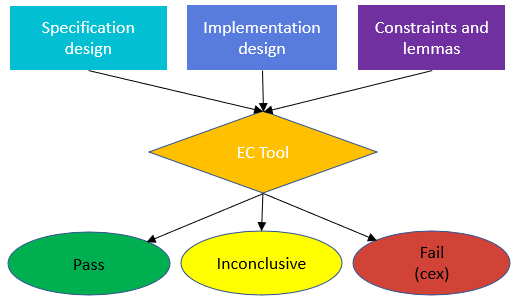}
      \caption{The inputs of an EC tool are two designs, specification and implementation, a set of constraints to drive the possible values to tests and a set of lemmas to prove. Each lemma can pass, fail or be inconclusive. A counterexample (cex) is provided for each failing lemma.}
  \label{fig:equivalence_checking}
  \end{figure}

  An alternative and successful approach is to rely on EC, defining two circuit representations to be equivalent if for all valid inputs they generate identical outputs. EC has been used in several contexts in the semiconductor industry~\cite{Drane2011FormalBlocks,Xue2013SimplificationModels}. The most popular types of EC are Boolean, Sequential and Transactional, and in this paper we focus on Transactional EC of combinational circuits, where the result of a given computation in the \emph{implementation} is compared against the result of the same computation in the trusted \emph{specification}. The output of the comparison can be \emph{pass}, when a property is proven, \emph{fail}, when the property is not true (a counterexample is generated), or \emph{inconclusive}, when the tool does not manage to either prove or disprove a property. See Figure \ref{fig:equivalence_checking}.
  
      
  A trusted reference is fundamental for EC. One standard verification flow used in the semiconductor industry is the following: starting from a component specification, a developer writes a high-level reference C++ design without any interaction with the designer who writes the RTL implementation, providing \emph{diversity} and \emph{independence} between the two, which are then formally tested for equivalence. Many more tests can be run on the C++ code, due to the great difference in simulation speed between C++ and RTL. This is usually described as \emph{C2RTL} EC. Another common option is what is called \emph{RTL2RTL} EC, where the reference is a trusted version of the same design in RTL, usually a version from previous projects or based on a third party library like Synopsys' DesignWare~\cite{Synopsys2021DesignS-2021.06-SP2}.

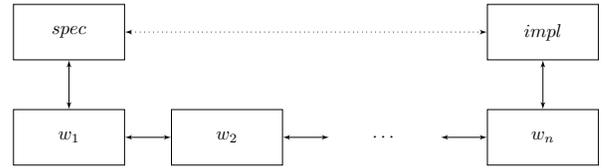
\begin{figure}
      \centering
      \scalebox{0.7}{
  
          \begin{tikzpicture}[auto, node distance=2cm,>=latex']
          \node [block] (spec) {$spec$};
          \node [block, right of=spec, node distance=9cm] (impl) {$impl$};
          \node [block, below of=spec] (w1) {$w_1$};
          \node [block, right of=w1, node distance=3cm] (w2) {$w_2$};
          \node [block, right of=w2, node distance=3cm, draw=white] (dots) { $\cdots$ };
          \node [block, below of=impl] (wn) {$w_n$};
      
          \draw [<->, dotted] (impl) -- node {} (spec);
          \draw [<->] (spec) -- node {} (w1);
          \draw [<->] (w1) -- node {} (w2);
          \draw [<->] (w2) -- node {} (dots);
          \draw [<->] (dots) -- node {} (wn);
          \draw [<->] (impl) -- node {} (wn);
      \end{tikzpicture}
      }
      \caption{Overview of the waterfall approach used by FV engineers. The dashed line between $spec$ and $impl$ represents an inconclusive verification. Full equivalence is achieved introducing $n$ intermediate designs $w_i$ and proving the equivalences of all the pairs ($spec$, $w_1$), ($w_1$, $w_2$), $\dots$, \mbox{($w_n$, $impl$)}. 
      }
      \label{fig:waterfall}
  \end{figure}

  
  Inconclusive results are commonplace in real-life EC and require advanced techniques to achieve full convergence, occupying most of the FV engineer's time. A common approach is to generate a ``\emph{waterfall}'', where the verification between implementation and specification is achieved by introducing intermediate designs, as shown in Figure \ref{fig:waterfall}. If all the intermediate equivalence steps are proven, the equivalence between specification and implementation holds.

  One of the key motivational papers for this work presents an overview of the technology behind Synopsys' industry leading Datapath Validation (DPV) tool~\cite{Koelbl2009SolverChecking}. The tool orchestrates a suite of techniques and solvers to prove the equivalence of input designs. One of these techniques is a set of rewrite engines. In this paper, the authors state that certain rewrite sets ``are only applied selectively'' or their application ``can be counter-productive''. As a result these rewrite engines are heuristic and may not explore the required space. The techniques presented in Section \ref{sec:proving_equiv} describe a rewrite orchestration approach that does not suffer from these limitations. 
  
  One relevant work combined rewriting and theorem proving to verify the correctness of gate-level multiplier designs in RTL~\cite{Temel2021SoundMultipliers,Temel2020AutomatedMultipliers}. In this work, the authors deploy ACL2 verified~\cite{Kaufmann1996ACL2:Nqthm} rewrites to transform optimized implementations into normalized implementations. Whilst our work targets a higher abstraction level, techniques and principles applied in the multiplier verification work will be relevant here. 
  
  \subsection{E-Graphs}
  E-graphs cluster equivalent expressions into e(quivalence)-classes, enabling a compact representation of alternative but functionally identical implementations. In the e-graph, nodes represent variables, constants or operators that point to children e-classes. This captures the intuition that we may choose how to implement a given sub-expression at any point in the design. Due to these nested choices, an e-graph can represent exponentially many implementations in the number of nodes.
  
\begin{figure}
      \centering
      \subfloat[Initial e-graph contains \newline $(2\times x)\gg 1$] {\includegraphics[scale=0.4]{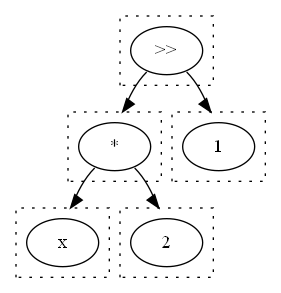}}
      \subfloat[Apply $x\times 2 \rightarrow x \ll 1$] {\includegraphics[scale=0.4]{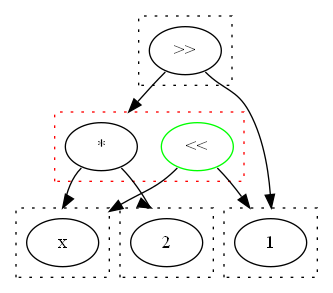}}
      \qquad
      \subfloat[Apply $(x\ll s)\gg s \rightarrow x$] {\includegraphics[scale=0.4]{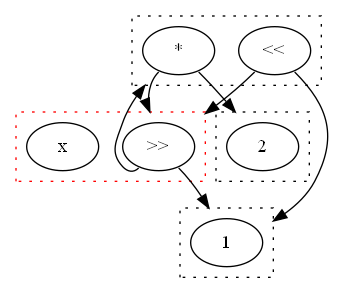}}
      \caption{Simple e-graph rewriting over the integers. The dashed boxes represent e-classes of expressions, where we have highlighted the modified e-class in red at each stage. Green nodes represent newly added nodes.
      }
      \label{fig: e-graph_example}
  \end{figure}

  An e-graph is grown via constructive application of local equivalence preserving rewrites, $l\rightarrow r$, where the right-hand side of the rewrite is added to the e-class containing $l$, without removing $l$ as in a traditional rewrite engine. As a result the e-graph avoids the phase-ordering problem, where the order of application impacts the results. This approach to growing an e-graph is known as equality saturation~\cite{Willsey2021Egg:Saturation,Tate2009EqualityOptimization,Joshi2002Denali:Superoptimizer}. A simple e-graph rewriting example is shown in Figure \ref{fig: e-graph_example}, where the dashed boxes represent e-class boundaries and the arrows connect nodes to their child e-classes. 
  
  The e-graph has been used in the formal methods community for many years~\cite{Nelson1980TechniquesVerification} and can be found in modern SMT solvers such as Z3~\cite{DeMoura2008Z3:Solver}. One particularly relevant work used e-graphs to perform translation validation of an LLVM compiler \cite{Stepp2011Equality-basedLLVM}. The recently released {\egg} library~\cite{Willsey2021Egg:Saturation} has enabled researchers to quickly develop a wide range of e-graph applications, ranging from hardware design~\cite{Coward2022AutomaticE-Graphs} to rewrite rule synthesis~\cite{Nandi2021RewriteSaturation}. By using e-graphs to represent datapath circuit designs in RTL we can take advantage of the e-graphs' ability to  explore equivalent designs efficiently. 
  
  \section{Proving Equivalence via E-Graph Rewriting} \label{sec:proving_equiv}
\begin{figure*}
      \centering
      \begin{tikzpicture}

\node [shape=rectangle,draw = black, text width = 2.3cm, text centered] at (-3,2) (spec) {Specification};
\node [shape=rectangle,draw = black, text width = 2.3cm, text centered] at (-3,-2) (impl) {Implementation};
\node [shape=rectangle,draw = black, text width = 1.2cm, text centered] at (-3,0) (slang) {Slang Parser};
\node [shape=rectangle,draw = red, minimum width=1.5cm, minimum height=1cm] at (0,0) (egraph) {E-graph};
\node [shape=rectangle,draw = red] at (0,1.2) (rewrites) {Rewrite};
\node [shape=rectangle,draw = red] at (0,-1.2) (analysis) {Analysis};
\node [shape=rectangle,draw = black, text width = 3cm, text centered] at (4,0) (extract) {Extraction \& Proof Production};


\node [shape=rectangle,draw = black, text centered] at (0,-2) (i1) {$I_1$};
\node [shape=rectangle,draw = black, text centered] at (2,-2) (i2) {$I_2$};
\node [shape=ellipse] at (4.5,-2) (dots) {$\cdot \cdot \cdot$};
\node [shape=rectangle,draw = black, text centered] at (7,-2) (im) {$I_m$};
\node [shape=rectangle,draw = black, text centered] at (7,-0.75) (istar) {$I^*$};

\node [shape=rectangle,draw = black, text centered] at (0,2) (s1) {$S_1$};
\node [shape=rectangle,draw = black, text centered] at (2,2) (s2) {$S_2$};
\node [shape=ellipse] at (4.5,2) (sdots) {$\cdot \cdot \cdot$};
\node [shape=rectangle,draw = black, text centered] at (7,2) (sn) {$S_n$};
\node [shape=rectangle,draw = black, text centered] at (7,0.75) (sstar) {$S^*$};

\draw [->,very thick] (spec) edge (slang);
\draw [->,very thick] (impl) edge (slang);
\draw [->,very thick] (-2.27,0.2) -> (-0.75,0.2);
\draw [->,very thick] (-2.27,-0.2) -> (-0.75,-0.2);
\path [->,very thick] (rewrites.east) edge[bend left] (egraph.north east);
\draw [-,very thick] (egraph.north west) edge[bend left] (rewrites.west);

\path [->,very thick] (analysis.west) edge[bend left] (egraph.south west);
\draw [-,very thick] (egraph.south east) edge[bend left] (analysis.east);

\draw [->,very thick] (egraph) edge (extract);


\draw [->,  thick, green!60] (extract) edge (i1);
\draw [->,  thick, green!60] (extract) edge (i2);
\draw [->,  thick, green!60] (extract) edge (im);
\draw [->,  thick, green!60] (extract) edge (istar);
\draw[<->,  thick] (impl) -> (i1) node[midway, below] {EC};
\draw[<->,  thick] (i1) -> (i2) node[midway, below] {EC};
\draw[<->,  thick] (im) -> (istar) node[midway, right] {EC};

\draw [->,  thick, green!60] (extract) edge (s1);
\draw [->,  thick, green!60] (extract) edge (s2);
\draw [->,  thick, green!60] (extract) edge (sn);
\draw [->,  thick, green!60] (extract) edge (sstar);
\draw[<->,  thick] (spec) -> (s1) node[midway, above] {EC};
\draw[<->,  thick] (s1) -> (s2) node[midway, above] {EC};
\draw[<->,  thick] (sn) -> (sstar) node[midway, right] {EC};

\draw[<->,  thick] (istar) -> (sstar) node[midway, right] {EC};

\end{tikzpicture}{}
      \caption{Flow diagram for the verification assistant, taking a specification and implementation circuit design in System Verilog. The designs are parsed and an e-graph is constructed. From the rewritten e-graph, extract two designs $S^*$ and $I^*$ along with intermediate designs forming a verification waterfall.}
      \label{fig:flow_diagram}
  \end{figure*}
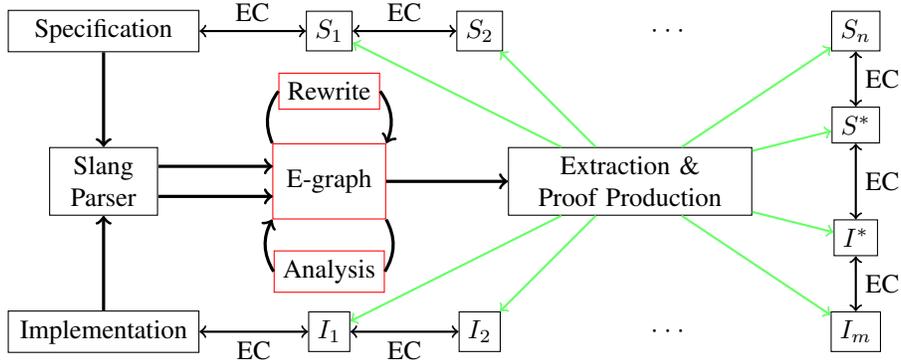

  The problem we tackle is, given specification and implementation RTL designs, prove them equivalent or reduce the original EC problem to a simpler one to solve. Given this objective, we will now describe how e-graph rewriting can provide an efficient solution. Figure \ref{fig:flow_diagram} illustrates the overall flow of the assistant. In this work we used a particular commercial tool throughout, but any RTL2RTL EC tool could be substituted in its place. 
  
  \subsection{E-Graph Initialization} \label{subsec:initialization}
  Using the framework developed in~\cite{Coward2022AutomaticE-Graphs}, we produce an e-graph representation of RTL, encoding all signal bitwidth and signage definitions. We use an intermediate language made up of nested S-expressions, as in Common Lisp \cite{Steele1990CommonLanguage}: 
  $$
  \texttt{term::=(operator [term] [term]\ldots [term])}
  $$
  For example, the following System Verilog:
  \begin{lstlisting}[language=Verilog]
      logic [7:0] a, b;
      logic [8:0] c;    
      always_comb c = a[7:0] + b[7:0];
  \end{lstlisting}
  corresponding to an unsigned addition of two primary 8-bit inputs \texttt{a} and \texttt{b}, stored in a 9-bit result is expressed as:
  $$
  \texttt{(+ 9 unsigned 8 unsigned a 8 unsigned b)}.
  $$
  The \texttt{+} operator takes eight arguments, describing the output and operand signals.
  
  This intermediate language is sufficient to correctly represent the functional behaviour of combinational RTL. Verilog operator definitions are context dependent meaning knowledge of all bitwidth and signage definitions is essential~\cite{Thomas2008TheLanguage}. In this work we target word-level RTL written in System Verilog. Using the open-source Slang parser~\cite{Popoloski2023Slang}, we implemented an automated flow converting System Verilog into this intermediate language. We parse both RTL designs and generate expressions, $S$ and $I$, in the intermediate language, for the specification and implementation respectively. 
  
  In most e-graph applications built using {\egg}, the e-graph is initialized with a single expression representing the design to be optimized. However, in this work we initialize the e-graph with both $S$ and $I$, such that the e-graph has two roots. The nodes common to both designs are automatically shared by {\egg}. In this paper we describe RTL generating a single output, but using constructs from \cite{Coward2022AutomaticE-Graphs} it is trivial to generalize to multiple outputs from each design. 
  
  \begin{figure}
      \centering
      \includegraphics[scale=0.35]{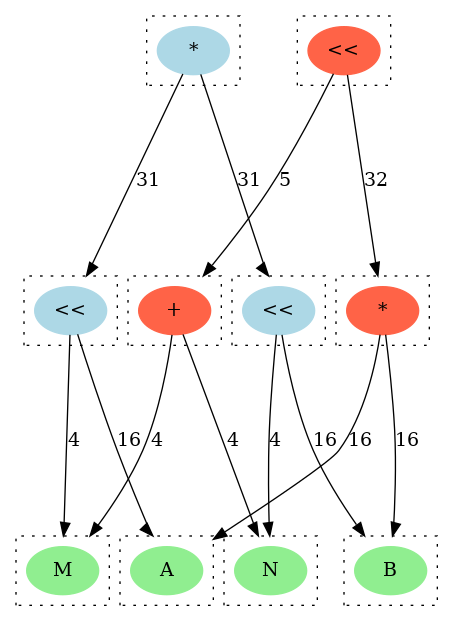}
      \caption{Initial e-graph representing two designs shown in Figure \ref{fig:motivational_verilog}, a specification (blue) and implementation (red). Shared nodes are colored green. Edge labels denote bitwidths. All e-classes (dashed boxes) initially contain a single node.}
      \label{fig:colored_egraph_0}
  \end{figure}
  
  In Figure~\ref{fig:colored_egraph_0}, we represent the two designs shown in Figure~\ref{fig:motivational_verilog} in a single e-graph. Colors indicate which design each node is used in. Note that, the designs initially only share the input variables and no intermediate signals. In the following sections we will discuss how as the e-graph is grown, common intermediate signals can be discovered. Initialising the e-graph with both designs means that we can simultaneously rewrite both designs in order to find a common equivalent.
  
  \subsection{Bitwidth Dependent Rewriting}
  The rewrites define the space of equivalent designs that can be reached as the e-graph grows. We build upon a subset of the bitwidth dependent rewrites described in~\cite{Coward2022AutomaticE-Graphs}, which was originally designed for optimization and was learnt from industrial RTL engineers. The optimization rewrite set deployed specific rewrites to improve correlation with the downstream logic synthesis tool. These rewrites are not deployed in the verification rewrite set. It is natural that the verification rewrite set should include many of the optimization capabilities but also incorporate additional verification specific rewrites that `undo' optimizations. For example, it may be productive to include transformations that introduce redundant logic that enables further sharing. 
  
  Table \ref{tab:rewrite_table} describes the small set of additional verification specific rewrites learnt from experience using commercial EC tools. Several of these rewrites are the reverse of rewrites targeting optimization. The space of rewrites that `undo' optimizations is less intuitive, so selecting valuable rewrites to include is challenging. We selected rewrites that were relevant for the test cases presented here. The assistant is designed such that it is simple for users to extend the rewrite set with their own transformations that are applicable to their designs.
  
  \begin{table}
      \caption{An example set of bitwidth dependent datapath verification rewrites. All rewrites are conditionally applied to ensure correctness. Bitwidth and signage information of operators and operands is omitted here for concision.}
      \centering
      \begin{tabular}{|c|c|c|}
           Name          & Left-hand Side & Right-hand Side \\
           \hline
           Unmerge Shift & $a\ll (b+c)$  & $(a \ll b) \ll c$               \\ 
           Mult Left Shift & $a \times (b\ll c)$  & $(a \times b) \ll c$   \\ 
           Shift to Mult & $a \ll \text{const}$ & $a\times 2^\text{const}$ \\ 
           Mult to Add   & $a \times 2$ & $a + a$ \\ 
      \end{tabular}
      
      \label{tab:rewrite_table}
  \end{table}
  
  One important consideration is to ensure that no rewriting opportunities are missed. To achieve this we parameterize the pattern matching left-hand side and apply the rewrites conditionally. We construct necessary and sufficient conditions that are functions of the rewrite parameters. Letting $l(\cdot)$ and $r(\cdot)$ denote functions mapping a vector of parameters $\vec{p}$, encoding operand bitwidth and signage, to expressions in the intermediate language. Given a parameterized rewrite, $l(\Vec{p}) \rightarrow r(\Vec{p})$, we construct a condition, $\phi$, such that $l(\Vec{p}) \cong r(\Vec{p}) \iff \phi(\vec{p})$. The sufficiency ensures that only valid, equivalence preserving, rewrites are applied. The necessity guarantees that no rewriting opportunities are missed. Missed opportunities can be the difference between a proven equivalence check and an inconclusive result. We will see this in Section \ref{sec:case_study}.
  
  A challenge for RTL level verification is that functional behaviour is bitwidth dependent, for example the addition of two 8-bit values stored in an 8-bit and a 9-bit result differ in general but may be equivalent under certain design constraints. We use the interval analysis and bitwidth reduction rewrites described in~\cite{Coward2023AutomatingE-Graphs}, deploying {\egg}'s built-in e-class analysis feature. These rewrites detect and reduce operators to the minimum bitwidth required to store the result, hence normalizing the operations. Such techniques are also deployed in commercial tools~\cite{Koelbl2009SolverChecking}, but program analysis on e-graphs is able to provide more precise abstractions~\cite{Coward2022AbstractE-Graphs}. 
  
  Having defined a set of rewrites, we use equality saturation to apply them to the e-graph initialized as described in Section \ref{subsec:initialization}. Rewrites are applied to both the specification and implementation designs simultaneously with the objective being to discover equivalent sub-expressions across the two designs. As rewrites are applied, new nodes are added to the e-graph and the e-classes grow, as we see in Figure~\ref{fig:egraphs}. We vary the number of e-graph rewriting iterations to control the e-graph growth throughout this work. Constructive rewrite application adds the overhead of maintaining many equivalent representations of the two designs in the e-graph, but greatly simplifies the problem of determining a correct rewrite application order.
  
  \subsection{Extraction} \label{subsec:extraction}
  Once the e-graph has saturated or reached a timeout, the e-graph represents two sets of equivalent designs, one for the specification and one for the implementation. From the e-graph we now seek to extract two designs, $S^* \cong S$ and $I^* \cong I$ that share the maximum number of common nodes. If $S$ and $I$ are found in the same e-class, namely the tool found a path of rewrites between the two designs, then $S^*$ and $I^*$ are identical. If they are found in different e-classes, we extract distinct $S^*$ and $I^*$ sharing as many of the common sub-expressions as is feasible from the e-graph. Figure \ref{fig:flow_diagram} shows the flow.
  
  To extract $S^*$ and $I^*$ we first identify which e-classes in the e-graph are associated with each design. Let $C$ denote the set of all e-classes. Given a root e-class, $r$, we recursively construct an associated $C_r \subseteq C$. Starting from $C_r = \emptyset$, we iterate through each node in $r$, adding its children e-classes to $C_r$, provided we had not already added it to $C_r$. We continue, recursively visiting each of the child e-classes and iterating through the contained nodes until $C_r$ stops growing. This construction is guaranteed to terminate. 
  
  Letting root(\textit{e}), be a function returning the root e-class for a given expression \textit{e} in the intermediate language, we use the algorithm described above to construct $C_\text{spec} \subseteq C$ starting from root($S$) and $C_\text{impl} \subseteq C$ starting from root($I$). We construct the shared e-class set, $C_\text{shared} = C_\text{spec} \cap C_\text{impl}$, which is used to identify the $S^*$ and $I^*$ that share the most common nodes. We update the spec and impl sets, $C'_\text{spec} =  C_\text{spec} \setminus C_\text{shared}$ and $C'_\text{impl} =  C_\text{impl} \setminus C_\text{shared}$. In Figures \ref{fig:colored_egraph_0} and \ref{fig:egraphs}, we highlight $C'_\text{spec}$ in blue,  $C'_\text{impl}$ in red and $C_\text{shared}$ in green. 
  
  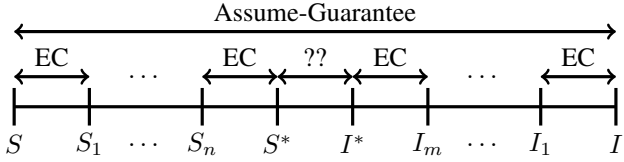
\begin{figure}
      \centering
      \begin{tikzpicture}

\node [] at (0,-0.5) (spec) {$S$};
\node [] at (1,-0.5) (spec) {$S_1$};
\node [] at (1.75,-0.5) (spec) {$\dots$};
\node [] at (1.75,0.4) (spec) {$\dots$};
\node [] at (2.5,-0.5) (spec) {$S_n$};
\node [] at (3.5,-0.5) (spec) {$S^*$};

\node [] at (4.5,-0.5) (impl) {$I^*$};
\node [] at (5.5,-0.5) (impl) {$I_m$};
\node [] at (6.25,-0.5) (impl) {$\dots$};
\node [] at (6.25,0.4) (impl) {$\dots$};
\node [] at (7,-0.5) (impl) {$I_1$};
\node [] at (8,-0.5) (impl) {$I$};

\draw [-,very thick] (0,0) edge (8,0);
\draw [-,very thick] (0,-0.25) edge (0,0.25);
\draw [<->,very thick] (0,0.4) -> (1,0.4) node[pos=0.5,above] {EC};
\draw [-,very thick] (1,-0.25) edge (1,0.25);
\draw [-,very thick] (2.5,-0.25) edge (2.5,0.25);
\draw [<->,very thick] (2.5,0.4) -> (3.5,0.4) node[pos=0.5,above] {EC};
\draw [-,very thick] (3.5,-0.25) edge (3.5,0.25);

\draw [<->,very thick] (3.5,0.4) -> (4.5,0.4) node[pos=0.5,above] {??};

\draw [-,very thick] (4.5,-0.25) edge (4.5,0.25);
\draw [<->,very thick] (4.5,0.4) -> (5.5,0.4) node[pos=0.5,above] {EC};
\draw [-,very thick] (5.5,-0.25) edge (5.5,0.25);
\draw [-,very thick] (7,-0.25) edge (7,0.25);
\draw [<->,very thick] (7,0.4) -> (8,0.4) node[pos=0.5,above] {EC};
\draw [-,very thick] (8,-0.25) edge (8,0.25);

\draw [<->,very thick] (0,1) -> (8,1) node[pos=0.5,above] {Assume-Guarantee};
\end{tikzpicture}
      \caption{E-graph extracted waterfall, generated automatically by the assistant. We use EC tools to prove the equivalence of the intermediate steps. The central equivalence check between $S^*$ and $I^*$, which may not be true, may not be provable using the EC tool, but represents a simplified problem.}
      \label{fig:intermediates}
  \end{figure}
  
  In previous work we have deployed hardware specific cost functions, for example circuit area or delay, that we seek to minimize in the extraction phase~\cite{Coward2022AutomaticE-Graphs,Coward2023AutomatingE-Graphs}. In this instance, we use a simpler objective function of e-graph nodes $n$:
  \begin{equation}
          \text{obj}(n)= 
  \begin{cases}
      K, & \text{if } \text{class}(n)\in C_\text{shared},\\
      -1,              & \text{otherwise},
  \end{cases}
  \end{equation}
  where $\text{class}(n)$ returns the e-class containing the node $n$ and $K$ is the total number of e-classes in the e-graph. We maximize this objective function to ensure that we share the maximum number of nodes possible, where the negative scoring of unshared nodes ensures that amongst designs sharing the same number of nodes, we extract the simplest one. We formulate the problem as an integer linear programming problem (ILP)~\cite{Wang2020SPORES:Algebra}. We define $N$ to be the set of nodes and $E \subseteq N \times C$ the set of edges. We also introduce $N_c$ to denote the set of nodes in a given e-class $c$ and $P_c$ to denote the set of parent nodes of $c$. For each node $n \in N$ we associate an objective, $\text{obj}(n)$, and a binary variable $x_n \in \{0,1\}$, which indicates whether $n$ is implemented in either of the extracted RTL designs. Lastly we introduce $R = \text{root}(I)\, \cup\, \text{root}(S)$. With these definitions the problem formulation is the following:
  \begin{align}
      \text{maximize}    \quad & \sum_{n \in N} \text{obj}(n) \cdot x_n  \label{eqn: max_obj}\\
      \text{subject to}  \quad & \forall (n,c) \in E\hspace{2.8em}: \; x_n \leq \sum_{\hat{n} \in N_c} x_{\hat{n}} \label{eqn: choose_child}\\
                         \quad & \forall c \in R\hspace{4.7em}: \; \sum_{n \in N_c} x_n = 1\label{eqn: all_outputs} \\
                         \quad & \forall c \in C\hspace{4.7em}: \; \sum_{n \in N_c} x_n \leq 1 \label{eqn: one_node_at_most}\\
                         \quad & \forall c \in C \text{ s.t. } P_c \neq \emptyset: \; \sum_{n \in N_c} x_n \leq \sum_{\hat{n} \in P_c} x_{\hat{n}}.\label{eqn: parents}
  \end{align}
  
  In the ILP problem, (\ref{eqn: choose_child}) guarantees that for every node $n$, we implement a node from each of its child e-classes, extracting only valid designs. (\ref{eqn: all_outputs}) then ensures that the outputs from both specification and implementation designs are produced by the extracted design. Lastly, \eqref{eqn: one_node_at_most} allows at most one node in each e-class to be implemented and \eqref{eqn: parents} ensures that only e-classes with implemented parents are selected, namely there are no unused signals in the generated RTL. We deploy topological sorting variables to handle cycles in the e-graph~\cite{Coward2022AutomaticE-Graphs,Wang2020SPORES:Algebra}. We use the Coin-Or CBC solver to solve the ILP problem. 
  
  For improved performance, we also use a comparable objective function that computes a greedy extraction based on \texttt{egg}'s built-in method. Such an approach is faster but fails to correctly account for common sub-expressions so may generate designs that are not as `close' as the ILP approach. We would recommend the ILP approach for solving EC problems that will require manual intervention.
  
  The extracted solution corresponds to two expression in the intermediate representation, $S^*$, equivalent to the specification and $I^*$, equivalent to the implementation, from which the tool automatically generates RTL. Using the recently added proof production feature in {\egg} \cite{Flatt2022SmallClosure}, two sequences of intermediate designs separated by a single rewrite are produced such that 
  \[S \cong S_1 \cong \dots \cong S_{n} \cong S^* \text{ and } I \cong I_1 \cong \dots \cong I_{m} \cong I^*.\]
  To remove the need to trust the correctness of the rewrites, the assistant generates System Verilog implementations for each of the intermediate designs and deploys the EC tool to formally verify the equivalence at each step as shown in Figure~\ref{fig:intermediates}. If the EC tool can prove each step including $S^* \cong I^*$, we have proven the equivalence of $S$ and $I$. Each intermediate proof is independent and can thus be proven in parallel. We can also specialize the solver configuration for each intermediate proof, since we are able to map rewrites to an optimal solver setup. For example, the commercial tool provides a set of solve scripts that handle proof orchestration with different capabilities. These scripts can be enabled by a user. We encoded a mapping from rewrites to the most efficient solve script in the assistant. With limited effort the assistant can be extended to target additional solvers.  
  
  To ensure soundness of the generated waterfall, a final ``Assume-Guarantee'' lemma proving $S \cong I$ is included, which uses all of the intermediate proofs (assuming they passed). This provides confidence that no gaps were left in the reasoning. If the tool is unable to prove $S^* \cong I^*$ then human intervention is required. However the EC problem is simplified, as these designs share more common signals than the original $S$ and $I$. 
  
  
  \section{Case-Study} \label{sec:case_study}
  \begin{figure*}
      \begin{subfigure}{\linewidth}
      \centering
      \includegraphics[scale=0.25]{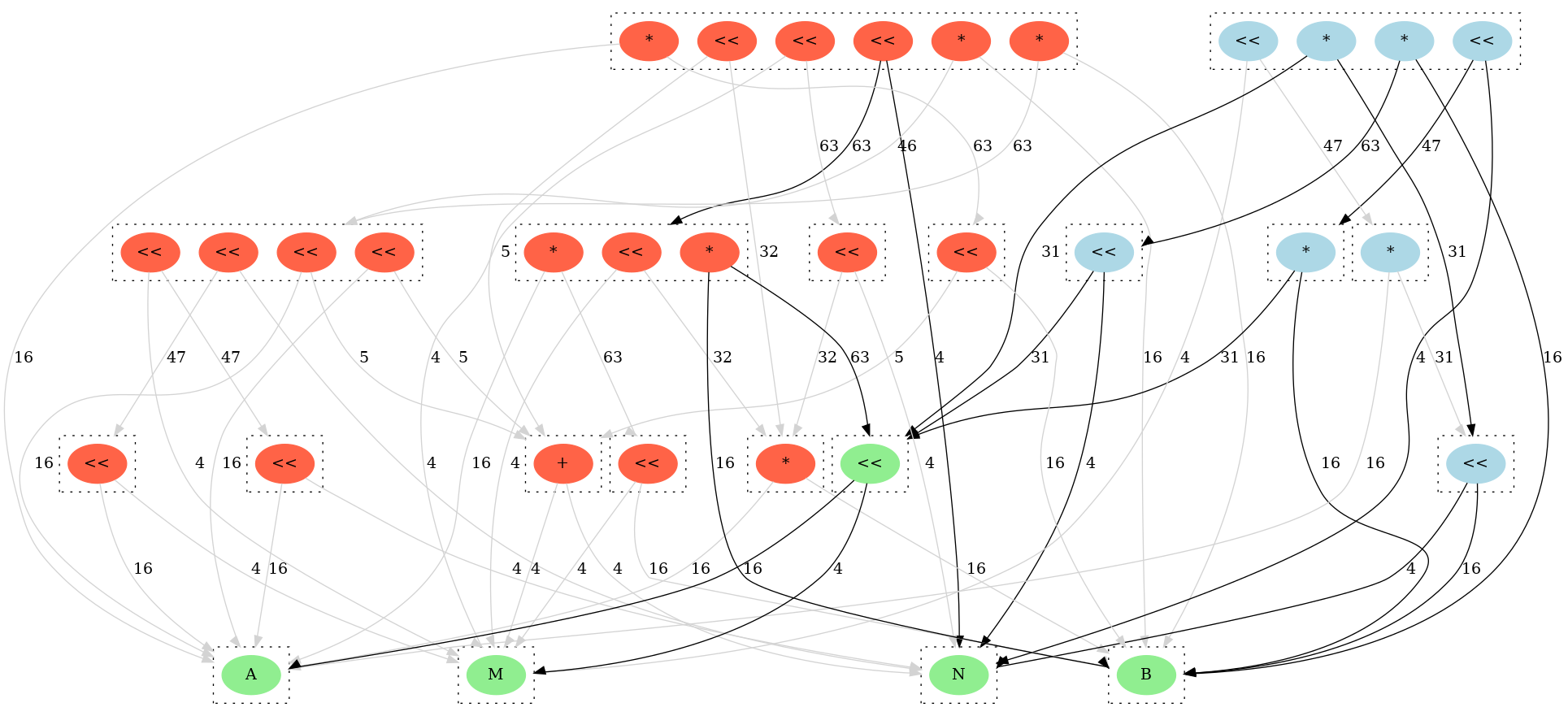}
      \caption{E-graph after two iterations of rewriting. Designs sharing an intermediate signal are highlighted with black arrows.}
      \label{fig:stage2}
      \end{subfigure}\quad
  \begin{subfigure}{\linewidth}
    \centering
      \includegraphics[scale=0.145]{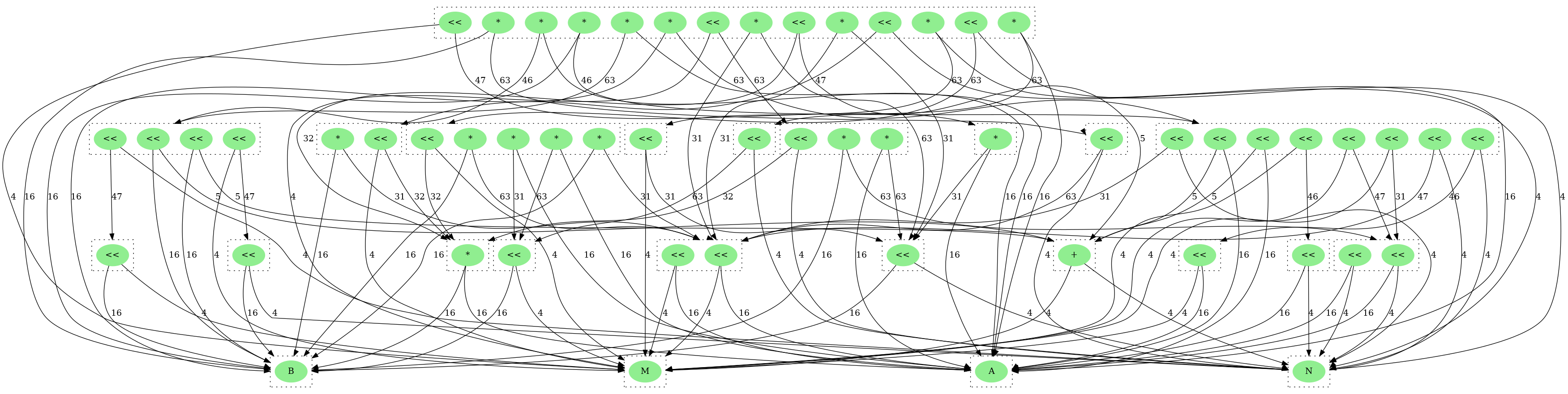}
      \caption{E-graph after three iterations of rewriting (77 nodes), where $S$ and $I$ have been merged into the same e-class.}
      \label{fig:stage5}
  \end{subfigure}
  \caption{Stages of e-graph growth starting from the initial e-graph in Figure \ref{fig:colored_egraph_0}.}
  \label{fig:egraphs}
  \end{figure*}
  
  We present a case-study of a real world problem where this technique proved beneficial. In all the following results we use an up to date version of the commercial EC tool running on SLES 12 on Intel Xeon W-2155 CPUs. 
  
  The designs shown in Figure \ref{fig:motivational_verilog}, are alternative ways to implement floating point multiplication of denormal numbers. More precisely, given two denormals  $2^{1-bias}\times 0.\text{mant}_a$ and $2^{1-bias}\times 0.\text{mant}_b$, the product of their mantissas is usually reduced to a standard non-denormal multiplication by shifting the values, expressing it as 
  either $(\text{mant}_a \ll m) \times (\text{mant}_b \ll n)$
  or equivalently as $(\text{mant}_a \times \text{mant}_b) \ll (m + n)$, 
  where $m = lzc(\text{mant}_a) + 1$, $n = lzc(\text{mant}_b) + 1$ and $lzc(\cdot)$ is the leading zero counter function.
  
  In three iterations of rewriting the e-graph applies a sequence of rewrites such that the specification and implementation are found within the same e-class. The progress of the e-graph can be seen in Figures \ref{fig:egraphs}. After two iterations of rewriting the first shared signal is detected, see the green left-shift in Figure \ref{fig:stage2}, where we have highlighted the initial specification and designs sharing the green node with brighter arrows. The e-graph shown in Figure \ref{fig:stage5}, after three iterations of rewriting, contains only green nodes, since the tool was able to apply a sequence of rewrites such that the original root nodes of $S$ and $I$ were merged into the same equivalence class. As a result, all e-classes are shared, meaning $C_{\text{shared}}=C$.
  
  From the final e-graph, Figure \ref{fig:stage5}, the tool then extracts identical $S^*$ and $I^*$ along with the sequence of rewrites that were applied to reach it. We summarise the rewrites applied below, omitting bitwidth alteration and commutativity steps.
  \begin{align}
      (A \times B) \ll (M + N)  &\rightarrow  \\
      \textit{Unmerge Left-Shift} \quad ((A \times B) \ll N) \ll M &\rightarrow \\
      \textit{Left-Shift Mult} \quad (A \times (B\ll N)) \ll M &\rightarrow \\
      \textit{Left-Shift Mult} \quad (A \ll M)\times(B\ll N) 
  \end{align}
  
  The e-graph assistant runs in 0.14 seconds, growing an e-graph comprised of 77 nodes. The EC tool is unable to prove the ``Left-Shift Mult'' and ``Mult Left-Shift'' transformations when non-uniform bitwidths are used. We resolve this by automatically inserting an additional intermediate step with standardized bitwidths. We hypothesize that this is due to a rewrite rule only being applied under certain parameterizations in the EC engine.
  
  Including all commutativity and bitwidth alteration rewrites, the assistant generated a total of 20 intermediate equivalence checks (including the ``Assume-Guarantee'' lemma) for the EC tool to prove. All intermediate proofs and the final completeness lemma are proven in 0.1 seconds by the EC tool. In contrast, when passed the original EC problem, $S\cong I$ with no assistance, the tool did not return a result within 24 hours. 
  
  \section{Results} \label{sec:results}
  
\begin{figure}
      \centering
      \includegraphics[scale=0.19]{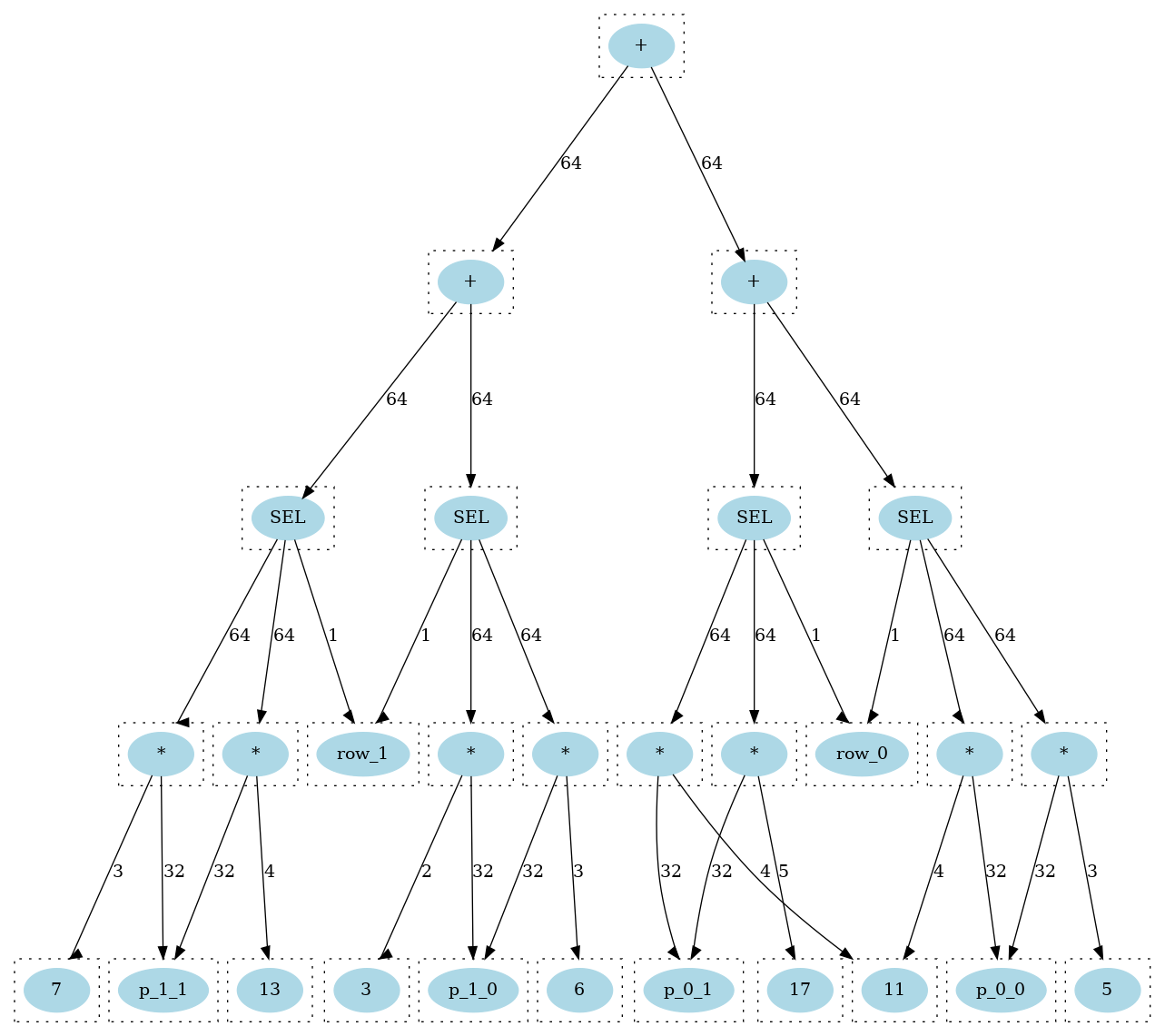}
      \caption{Dataflow graph of the initial box filter design. The \texttt{SEL} nodes represent muxes.}
      \label{fig:box_filter_dfg}
  \end{figure}

  \begin{table*}
    \centering
    \caption{Industrial EC tool performance with and without intermediate proofs generated by the assistant. We report the baseline EC tool performance when solving the original EC problem. We also report the runtime of the e-graph assistant and the runtime of the EC tool when solving the problem with the intermediate proofs. The sum provides a total verification time for the assisted proof. The last column shows the speedup ratio achieved using the assistant. Runtimes are in seconds.}
    \label{tab:results_table}
    \begin{tabular}{|l|r|r|r|r|r|}
    \hline
         Benchmark & EC without assistance & Assistant & EC with assistance & Assisted Total & Speedup (without/with)\\
         \hline
         ADPCM Decoder & \textbf{0.68} & 0.38 & 0.49 & 0.87 & 0.78 \\
         H-264 VBSME-4 & 7.93 & 7.04 & 0.71 & \textbf{7.75} & 1.02\\
         H-264 VBSME-8 & 93.13 & 14.3  & 0.20 & \textbf{14.50} & 6.42 \\
         FIR Filter    & 5.50 & 3.49  &  0.79 & \textbf{4.28}  & 1.29\\
         Box Filter    & 79.56 & 16.10 & 1.61 & \textbf{17.71} & 4.49 \\
         Case Study    & - & 0.14  & 0.10 & \textbf{0.24} & - \\ 
         \hline
    \end{tabular}

\end{table*}
  Having demonstrated how the verification assistant can provide the intermediate steps transforming a previously inconclusive proof into one solved in under one second, we will now demonstrate how the assistant can improve overall verification runtimes across a datapath optimization benchmark set. We take benchmarks from \cite{dataflow2008verma} and implement original and optimized RTL for those designs that are fully described in this paper. The ADPCM decoder is an approximate multiplication implementation. We include two instances of a kernel from the H.264 VBSME (variable block size motion estimator), which correspond to absolute difference summation trees of size four and eight, $\sum_i |a_i-b_i|$. The FIR Filter is a typical finite impulse response filter of depth eight. The case study and box filter are Intel provided benchmarks. The box filter is a reconfigurable square filter, sampling four pixels at a time. The dataflow graph for this design is shown in Figure~\ref{fig:box_filter_dfg}. The optimized design deploys constant factorization and mux rewriting which is relatively challenging for the EC tool to prove.
  
  The benchmarks include a range of arithmetic and logical operators, representative of typical RTL optimizations that may be performed by hand or by a specialized datapath optimization tool. For each benchmark, we run the assistant until either, it discovers a complete path between specification and implementation or it deploys five iterations of rewriting, whichever comes first. The e-graph applies all rewrites in parallel at each iteration, meaning that many parts of the designs can be simultaneously transformed in each iteration. 
  
  In these results, the EC tool does not report any increase in the initial compilation time, which is less than a second for all cases presented here. We report the runtime from when the solvers start running. For the baseline, we deploy all the EC tool's solvers in parallel and take the fastest proof. When the verification assistant has generated a sequence of intermediate proofs, we report the maximum time taken to solve a single sub-problem, since each proof can be run in parallel. In practice, the industrial tool's multi-processor environment introduced runtime overhead that was not related to the proof. Namely, running a proof on a server grid produced unpredictable runtime results due to the licence checks and interactions with the workload management software.
  
  Table \ref{tab:results_table} presents the performance impact of the assistant on the total verification time. In the first example, ADPCM Decoder, the EC tool efficiently proves the correctness of the two designs, meaning that the overhead of the assistant is detrimental, increasing total runtime. It is worth noting that the intermediate proofs do help reduce the solve time. 
  
  In the remaining benchmarks, the baseline EC tool takes longer to prove equivalence. The introduction of intermediate proofs reduces the EC solve time by up to 465x, when we just compare the EC tool runtimes and discount the assistant's runtime. Including the runtime to generate the intermediate proofs, the total verification time is reduced by up to 6x. In most cases, the EC tool solves each of the intermediate proofs in less than 0.5 seconds as each step represents a single local modification to the design. The assistant can effectively select the most optimal solver orchestration script per intermediate proof, which greatly helps performance. This is possible because the assistant understands what transformation has been applied at each stage. Such an approach avoids wasted compute resources, since there is no need to run different solvers in parallel for each of the intermediate problems.
  
  \begin{table}
    \centering
    \caption{Summary of e-graph assistant properties across the benchmarks. We report the number of rewriting iterations, the e-graph size in terms of node count, the number of intermediate proofs generated, and whether the e-graph found a complete path of rewrites between $S$ and $I$.}
    \label{tab:egraph_table}
    \begin{tabular}{|l|r|r|r|c|}
    \hline
         Benchmark & Num. Iter. & E-graph Nodes & Num. Proofs & Full Path \\
         \hline
         ADPCM         & 3 & 469 & 20 & Y \\
         VBSME-4       & 5 & 5640 & 26 & Y\\
         VBSME-8       & 5 & 5800 & 46 & Y\\
         FIR Filter    & 5 & 4700 & 23 & Y\\
         Box Filter    & 5 & 21400 & 115 & N\\
         Case Study    & 3 & 149 & 20 & Y\\ 
         \hline
    \end{tabular}

\end{table}
  
  
  
  The box filter is an Intel provided benchmark and is the only example where the assistant is unable to find a complete path. This verification problem may require additional rewriting iterations or entirely new rewrites to reach the implementation design. To minimize runtime, we deploy the faster greedy extraction method. To solve the  EC problem, $S^* \cong I^*$, we default to one of the slower but more powerful solver orchestration scripts. In this case, the $S^* \cong I^*$ EC problem takes significantly longer to prove than the other sub-problems. In general, as the assistant is able to deploy longer sequences of dependent rewrites, corresponding to more iterations of rewriting, we expect to find $S^*$ and $I^*$ that are increasingly close. In Table \ref{tab:results_table}, we reported box filter results based on five iterations of rewriting. If we instead limit the e-graph to three iterations of rewriting, the assistant's runtime is reduced from 16 seconds to 2 seconds. The intermediate proofs generated by this smaller e-graph can be proven in 1.12 seconds, reducing the total verification time to approximately 3 seconds, corresponding to a 24x speedup over the baseline. 
  
  The box filter results highlight a tradeoff between resource investment into generating intermediate proofs and into solving these proofs. Understanding the turning point would allow the assistant to automatically identify which intermediates will be most beneficial, a task beyond the current tool. 
  
  In addition to the solvers described so far, we also investigated the open-source SymbiYosys equivalence checker~\cite{YosysHQGmbHSymbiYosys}. However, for the instances we tried, we were not able to solve any equivalence problems since the tool is SAT/SMT based and does not handle datapath problems efficiently. A key advantage of a rewrite based approach is that performance is not affected by bitwidths, whilst SAT-based solvers will suffer from exponential slowdowns as the bitwidths are increased. This approach is promising because we do not typically need the full power of a SAT or SMT solver on the entire design, meaning that a specialized tool that does not target notions of completeness can prove valuable.
  \section{Conclusion}
  This paper applies recent advances in e-graph rewriting to datapath equivalence checking to develop an automated formal verification assistant that enhances the capabilities of industrial tools. By incorporating both the specification and implementation into a single data structure, the assistant simultaneously rewrites both designs to efficiently identify common equivalent sub-expressions. From the e-graph, the tool extracts a sequence of intermediate designs, breaking the complete equivalence check into a sequence of smaller sub-proofs which can be proven by \emph{trusted} tools. In cases where the assistant is unable to identify a complete path between the specification and the implementation designs, the e-graph rewriting may still reduce the equivalence checking to a simpler sub-problem. This enables FV engineers to focus on the challenging core of the verification task and helps the EC tool to identify additional internal equivalence pairs automatically, reducing the complexity of the overall equivalence check.
  

  The assistant developed through this work is able to find a complete sequence of intermediate designs, enabling a commercial EC tool to prove equivalence in under a second on a problem that was previously beyond its capabilities. We also demonstrated test cases where the verification assistant was able to reduce verification runtimes by up to 6x.
  
  Future work will primarily investigate integration of the techniques presented in this paper into complete solvers to improve the rewrite engines in such tools. We will also explore different front-ends to enable C to RTL equivalence checking and will incorporate registers to facilitate equivalence checking across multiple cycles. Exploring alternative applications, such as the gate-level multiplier verification challenge discussed in the background, would highlight the generality of the approach. Lastly, there are many performance optimizations that we will make to the assistant. For example, having discovered shared classes in the e-graph, we could freeze these sub-graphs to limit e-graph growth. Such optimizations and better orchestration would allow us to extend the evaluation to larger inconclusive problems requiring deeper e-graph exploration.

  \bibliographystyle{IEEEtran}
  \bibliography{IEEEabrv,references2}
  
  \end{document}